# Optimizing Superconducting Nb Film Cavities by Mitigating Medium-Field Q-Slope Through Annealing

B. Abdisatarov [1,2], G. Eremeev [2], H. E. Elsayed-Ali [1], D. Bafia [2], A. Murthy [2], Z. Sung [2], A. Netepenko [2], A. Romanenko [2], C. P. A. Carlos [3], G. J. Rosaz [3], S. Calatroni [3], S. Leith [4], A. Grassellino [2]

[1] Department of Electrical and Computer Engineering, Old Dominion University, Norfolk, Virginia 23529, USA and Applied Research Center, 12050 Jefferson Avenue, Newport News, Virginia 23606, USA

[2] Fermi National Accelerator Laboratory, Batavia, Illinois 605010, USA

[3] CERN, European Organization for Nuclear Research, 1211 Geneva, Switzerland

[4] IMS Nanofabrication GmbH, Dresdner Str. 47 1200 Vienna, Austria

# ABSTRACT

Niobium films are of interest in applications in various superconducting devices, such as superconducting radiofrequency cavities for particle accelerators and superconducting qubits for quantum computing. In this study, we address the persistent medium-field Q-slope issue in Nb film cavities, which, despite their high-quality factor at low RF fields, exhibit a significant Q-slope at medium RF fields compared to bulk Nb cavities. Traditional heat treatments, effective in reducing surface resistance and mitigating the Q-slope in bulk Nb cavities, are challenging for Nb-coated copper cavities. To overcome this challenge, we employed DC bias high-power impulse magnetron sputtering to deposit Nb film onto a 1.3 GHz single-cell elliptical bulk Nb cavity, followed by annealing treatments aimed at modifying the properties of the Nb film. In-situ annealing at 340 °C increased the quench field from 10.0 to 12.5 MV/m. Vacuum furnace annealing at 600 °C and 800 °C for 3 hours resulted in a quench field increase of 13.5 and 15.3 MV/m, respectively. Further annealing at 800 °C for 6 hours boosted the quench field to 17.5 MV/m. Additionally, the annealing treatments significantly reduced the field dependence of the surface resistance. However, increasing the annealing temperature to 900 °C induced a Q-switch phenomenon in the cavity. The analysis of RF performance and material characterization before and after annealing has provided critical insights into how the microstructure and impurity levels in Nb films influence the evolution of the Q-slope in Nb film cavities. Our findings highlight the significant roles of hydrides, high local misorientation, and lattice and surface defects in driving field-dependent losses. By strategically optimizing film properties and controlling impurity levels, we demonstrate a promising pathway to mitigate the medium-field Q-slope, paving the way for more efficient superconducting RF technologies.

## I. INTRODUCTION

Superconducting radiofrequency (SRF) cavities have emerged as indispensable components in modern particle accelerators, offering cost-effective solutions for large-scale accelerators. SRF cavities are typically made from Nb metal sheets via stamping and welding. The understanding of the chemical and physical properties of bulk Nb material has been advanced to tailor cavity performance to accelerator requirements [1–3]. Instead of using expensive Nb metal, coating Nb film onto inexpensive substrates such as Cu offers a cost-effective solution, particularly for large low-frequency cavities. High thermal conductivity and lower sensitivity to DC magnetic field are other benefits enabled by Cu substrates. Nb film on Cu cavity (Nb/Cu) technology has been successfully developed and implemented for the Large Electron-Positron Collider energy upgrade (LEP-II), the Large Hadron Collider (LHC), and, more recently, deployed onto SRF cavities for HIE-ISOLDE facility [4–8]. Future Circular Collider (FCC), which is the proposed upgrade to LHC, will require a large number of Nb/Cu cavities operating at higher accelerating gradients [9]. Efforts have been ongoing to advance Nb/Cu technology for the proposed FCC project [10].

Past studies have demonstrated that, while a single-cell TESLA shape elliptical Nb/Cu cavities exhibit high intrinsic quality factor ($Q_0$) at low RF fields (< 4 MV/m), they encounter stronger Q-slope at medium RF fields (4 - 20 MV/m), unlike bulk Nb cavities with high residual resistivity ratio (RRR) [1, 2]. Several mechanisms have been proposed to explain anomalous Q-slope in Nb film cavities. Early flux penetration due to shorter electron mean free path [11], superconducting gap suppression [12], Nb film granularity [7, 13] and surface roughness [14], impurities [15], film microstructure and the Nb/Cu interface [16, 17], and substrate surface defects [18] are among hypotheses proposed over the years. To explore and mitigate these effects, different Nb film coating techniques have been explored since the 1980s. Besides DC magnetron sputtering, DC-biased cathode sputtering [19], vacuum arc deposition [20], coaxial energetic deposition [21, 22], electron cyclotron resonance plasma deposition [23, 24], and high power impulse magnetron sputtering (HiPIMS) [25] are the Nb film deposition techniques which have been applied to SRF cavities. Each of these techniques has advantages and disadvantages in their application to SRF cavity coating. In recent investigations, HiPIMS coating has emerged as the most promising technique for coating Nb superconducting films without a strong Q-slope [25–28]. This technique enhances film characteristics through a higher ionization ratio, increased ion energy, and controlled the energy of the film's species. As a result, HiPIMS-deposited Nb films exhibit excellent adhesion, low roughness, and a dense columnar structure, often resulting in a better surface resistance ($R_s$) [28].

In the development of Nb film coating techniques for SRF applications, a range of parameters have been explored and optimized [1, 2]. While post-processing has been significantly improved for bulk Nb SRF cavities over the past several decades, it remains less studied for Nb/Cu SRF cavities, largely due to the challenges posed by the thermal and chemical properties of Cu. Heat treatments at 350 °C were investigated as a method to reduce hydrogen content in Nb films deposited on Cu; however, no significant reduction was observed [29]. Annealing at 120 °C—commonly used for bulk Nb cavities—was also found to be detrimental to Nb films [30]. Although heat treatments over a wide temperature range (120 to 950 °C) have proven effective in reducing or even reversing the Q-slope in bulk Nb cavities [31–33], such treatments are difficult to apply to Nb/Cu systems due to the lower softening temperature of Cu and the potential for interdiffusion at the Nb/Cu interface.

To overcome these limitations and allow for a systematic investigation of post-deposition treatments, we adopted an alternative approach by depositing Nb films onto bulk Nb substrates rather than Cu. This substitution enables high-temperature annealing without the thermal constraints of Cu, allowing us to isolate and study the role of Nb film impurities, lattice strain, and microstructural evolution—factors that are believed to contribute to performance limitations such as the medium-field Q-slope in Nb/Cu cavities.

Researchers have extensively investigated Nb films deposited on Nb and Cu substrates [5, 15, 27, 28, 29]. Based on their studies and our material characterization, we found that Nb films deposited on both Nb and Cu substrates exhibit very similar characteristics in terms of film growth, surface roughness, grain size, impurity levels, RRR, and microstructure.

In this work, we demonstrate that medium and high-temperature annealing of Nb films on Nb substrates mitigates the medium-field Q-slope by reducing impurity content and relieving microstrain. These insights provide valuable guidance for future efforts to improve Nb/Cu coatings by focusing on hydrogen reduction, strain relaxation, and surface defect optimization—objectives that can ultimately enhance RF performance in practical Nb/Cu SRF cavities.

## II. EXPERIMENT

A single-cell TESLA shape elliptical 1.3 GHz Nb cavity was used for film deposition. The fine grain (∼ 50 μm) bulk Nb cavity with a RRR of 300 underwent the standard preparation steps in accordance with established practices [1]. First, the cavity was electropolished to remove about 120 μm. Subsequently, 3 hours annealing at 800 °C in a vacuum furnace was applied. Then, a light electropolishing step removed about 25 μm from the surface. Ultrasonic cleaning in a heated Liquinox solution, high pressure rinsing with ultra-pure water, and assembly with vacuum and RF hardware in ISO4 cleanroom were conducted for the final preparation before cryogenic testing. The $Q_0$ as a function of the accelerating gradient ($E_{acc}$) was measured at 2.0 and 1.5 K for the baseline testing in a liquid helium dewar at Fermilab's vertical cavity testing facility (VCTF) [34].

A DC-biased HiPIMS deposition system developed at CERN was used to coat the inner surface of the bulk Nb SRF cavity with a 6 μm-thick Nb film. Prior to deposition, the vacuum chamber was baked out to reach a base pressure of approximately $10^{-10}$ mbar removing residual gases, oxides, and contaminants. For deposition, high-purity krypton gas (99.998%) was introduced in the chamber at a pressure to ∼3 × $10^{-3}$ mbar. The deposition was carried out at 150 °C using a cylindrical Nb target with a RRR of ∼300. A detailed description of the HiPIMS system and deposition parameters can be found in [27, 28]. In parallel, small Nb samples (10 × 10 × 3 mm) were prepared and coated using the same cleaning, deposition, and post-processing conditions as those applied to the cavity. These samples were used for material characterization to provide a direct correlation with the RF performance of the coated cavity.

A Thermo Fisher Scientific Helios 5 DualBeam scanning electron microscope and focused ion beam (SEM-FIB) system was utilized to analyze the surface morphology and film thickness. Measurements were conducted on randomly selected regions at an accelerating voltage of 5 kV and a current of 1.4 nA. A gallium-focused ion beam was employed for milling to obtain cross-sectional and electron-transparent images of the film. For milling, a 1 μm thick platinum layer was initially deposited to protect the surface. Surface plane-view and transmission (T-) electron backscatter diffraction (EBSD) mapping were performed. Surface EBSD maps were used to observe grain size, crystallographic orientation, and local misorientation angles, and T-EBSD allowed to observe smaller grains less than 100 nm in diameter size. NanoMagnetics Instruments atomic force microscope (AFM) was employed for surface topology to examine randomly selected 10 × 10 μm areas and measure the root mean square (RMS) surface roughness. Surface chemistry and bulk structural properties of the Nb film were determined by a systematic study using time of flight secondary ion mass spectrometry (ToF-SIMS), X-ray photoelectron spectroscopy (XPS), X-ray diffraction (XRD), and transmission electron microscopy (TEM) as previously reported [35].

RF performance of the Nb film cavity was tested in the VCTF. The thermometry mapping (TMAP) system was used to monitor the distribution of power dissipation on the cavity surface. TMAP makes it possible to identify the location of quench, which refers to the sudden loss of superconductivity of the cavity [36]. The TMAP system consists of 36 boards evenly positioned at 10° intervals around the surface of the cavity. Each board is equipped with 16 resistive temperature sensors, resulting in a total of 576 sensors on the cavity surface. To prevent the superfluid helium from directly contacting the sensor surface during measurement, Apiezon cryogenic grease is applied to the face of each resistor. The TMAP measurement efficiency is approximately 35% which is the measure of how much the temperature rise at the cavity's outer surface compares to the temperature rise at the inner surface [36]. This efficiency indicates the system's capability to quantify surface heating. Temperature mapping plays a crucial role in the investigation and analysis of losses in SRF cavities, aiding in the understanding and improvement of their performance.

A series of annealing treatments was applied to the Nb film coated cavity. Initially, an *in-situ* annealing at 340 °C for 1 hour was conducted to assess the impact of the native oxide layer on the film surface. This treatment utilized specialized oven and vacuum hardware to maintain vacuum within the cavity throughout the process. As a result, the oxide layer was dissolved, and the cavity was subsequently tested without exposure to air. Following this, four additional annealing treatments were carried out in a conventional vacuum furnace typically used for hydrogen degassing of SRF cavities: 3 hours at 600 °C, 3 hours at 800 °C, 6 hours at 800 °C, and 6 hours at 900 °C. After each of these furnace annealing steps, the SRF cavity was exposed to the air and the cavity was subjected to high-pressure rinsing (HPR) to restore surface cleanliness. As expected, a native oxide layer reformed on the film surface due to air exposure. RF measurements were conducted in the VCTF following each annealing step to evaluate changes in performance.

## III. RESULTS

### A. Nb film characterization

SEM images of the Nb films revealed a textured surface morphology without any cracks or droplets, and the cross-sectional analysis indicated an average thickness of 6.5 ± 0.5 µm for the Nb film, as shown in Figure 1 (a) and (b). AFM images confirmed a relatively smooth surface with the RMS surface roughness in randomly selected 10 × 10 µm areas determined to be 13.4 ± 2 nm, as shown in Figure 1 (c). All other detailed material analyses can be found in Abdisatarov et al. [35].

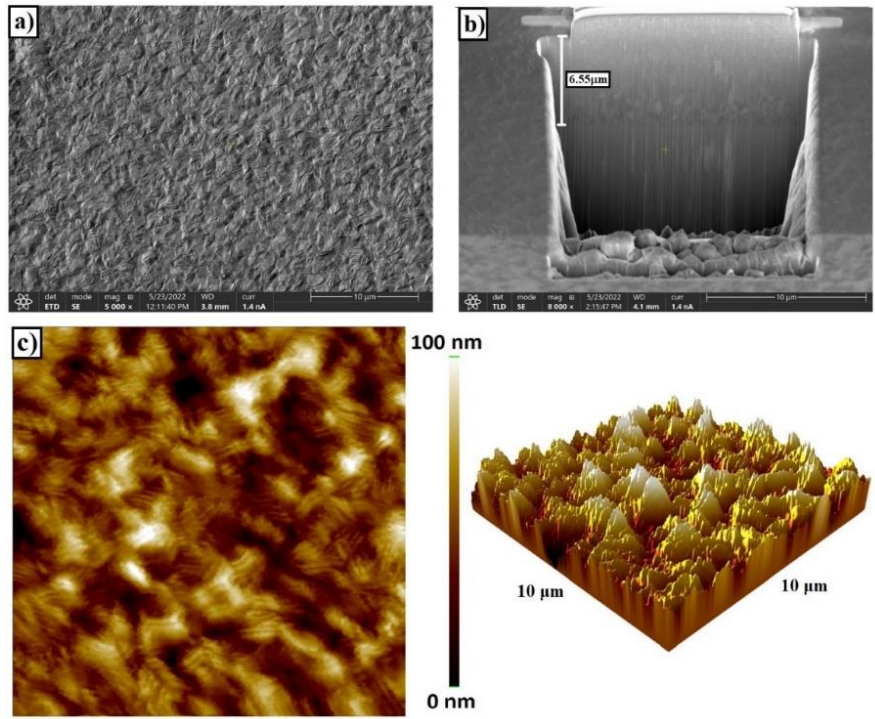

Figure 1. SEM and AFM images of Nb film on Nb substrate (a) surface morphology, (b) cross sectional SEM image of Nb film, (c) 2D and 3D AFM images of 10 × 10 µm surface area. The RMS surface roughness in the AFM-imaged area is 13.4 ± 2 nm.

### B. RF results

The baseline testing of the bulk Nb cavity prior to film deposition demonstrated the expected field dependence typical of an electropolished cavity [2]. The cavity was limited by high-field Q-slope at $E_{acc} \cong 30$ MV/m, as illustrated in Figure 2. Following the film deposition, a notable degradation in $Q_0$ was observed and the cavity exhibited a pronounced Q-slope. The maximum attainable field was constrained by a quench at $E_{acc} \cong 10.0$ MV/m, as indicated in Figure 2. Despite *in-situ* annealing for 1 hour at 340 °C, which aimed to mitigate the field dependency, the cavity maintained a strong Q-slope, though the quench field improved by ~ 25%, reaching $E_{acc} \cong 12.5$ MV/m. A subsequent treatment involving vacuum furnace annealing at 600 °C for 3 hours led to a noticeable reduction in the field dependence of the $R_{res}$. Nevertheless, the quench field showed only a marginal increase, reaching $E_{acc} \cong 13.5$ MV/m, which is within the error margin of the previous treatment. Conducting vacuum furnace annealing for 3 hours and 6 hours at 800 °C revealed additional reductions in $R_s$. Consequently, significant improvements in $Q_0$ were recorded, with the maximum field increasing to $E_{acc} \cong 15.3$ MV/m and eventually reaching $E_{acc} \cong 17.5$ MV/m. After subjecting the cavity to vacuum furnace annealing at 900 °C for 6 hours, no significant chnage in the Q-slope was observed. However, a new phenomenon emerged known as the Q-switch [1]. This effect was characterized by a degradation in performance occurring at an accelerating field gradient of $E_{acc} \cong 11.0$ MV/m, which subsequently led to a quench at $E_{acc} \cong 12.3$ MV/m, as shown in Figure 2.

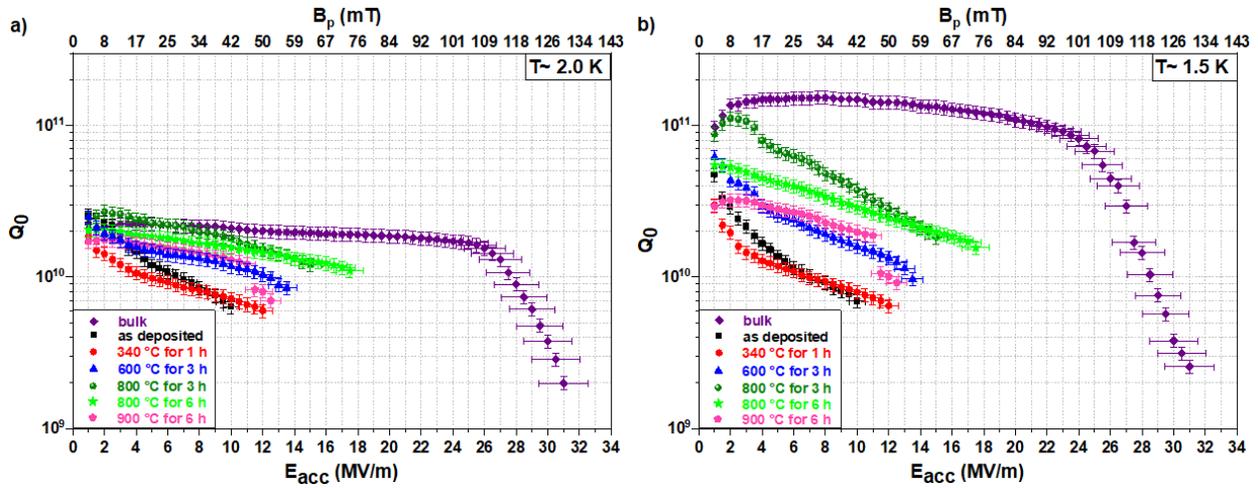

Figure 2. $Q_0$ versus $E_{acc}$ at T = 2.0 K (a) and 1.5 K (b) for the bulk Nb cavity and Nb film cavity before and after annealing at different temperatures and durations. The Nb film cavity demonstrated a field-dependent $Q_0$, while the heat treatments effectively reduced this field dependence and enhanced the critical quench field.

The $R_s$ was calculated using the fundamental relation between the $Q_0$ and the geometry factor (G). Specifically, is given by the ratio $G/Q_0$: where the $G = 270$ for this cavity [35]. Figure 3 presents the decomposition analysis of $R_s$ into two components: the residual resistance ($R_{res}$) and the BCS resistance ($R_{BCS}$) for each VCTF test. At 1.5 K, the contribution of $R_{BCS}$ is negligible compared to $R_{res}$, allowing us to approximate $R_s \approx R_{res}$ at this temperature. Given that $R_{res}$ is temperature-independent in the superconducting regime, the value of $R_{BCS}$ at 2.0 K can then be extracted using the relation $R_s(2.0\ K) = R_{res}(2.0\ K) + R_{BCS}(2.0\ K) \approx R_s(1.5\ K) + R_{BCS}(2.0\ K)$ [37, 38]. In the baseline test of the bulk Nb cavity, both $R_{res}$ and $R_{BCS}$ exhibited weak dependence on the applied field, remaining stable from an accelerating field gradient of $E_{acc} \cong 2$ MV/m to $E_{acc} \cong 18$ MV/m as anticipated. Following the deposition of the Nb film, $R_{BCS}$ decreased significantly from approximately 10-12 n$\Omega$ in the baseline test to about 2-3 n$\Omega$. In contrast, $R_{res}$ demonstrated a strong field dependence post-deposition, increasing by more than one order of magnitude compared to the bulk Nb. *In-situ* annealing at 340 °C resulted in a substantial increase in $R_{res}$, while $R_{BCS}$ remained relatively unchanged. Subsequent higher temperature vacuum furnace annealings elevated $R_{BCS}$ to approximately 6-8 n$\Omega$, while also significantly reducing both the magnitude and the field dependence of $R_{res}$, as illustrated in Figure 3.

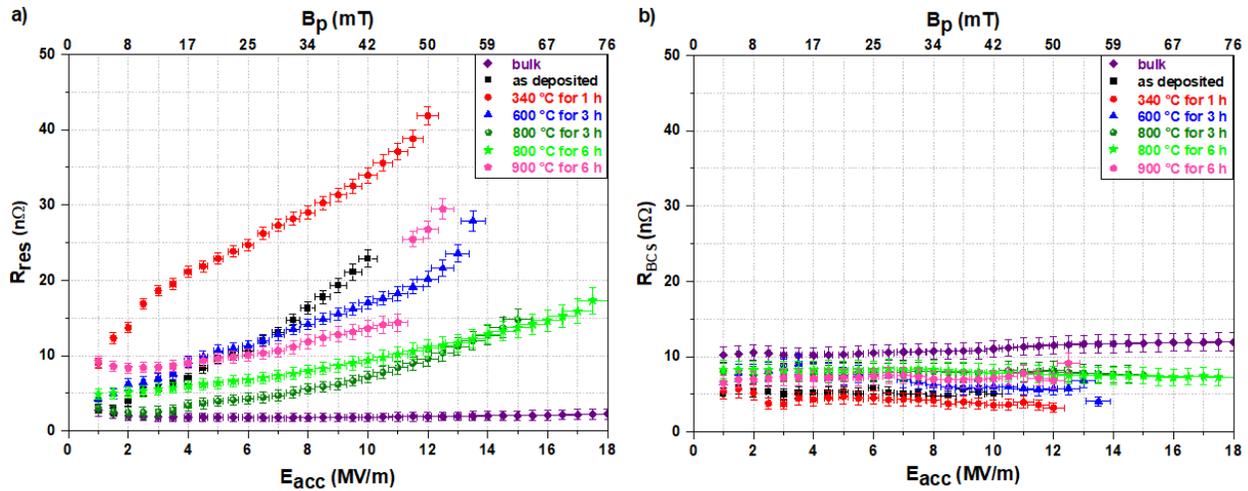

Figure 3. The decomposition analysis of the $R_s$ into the $R_{res}$ (a) and $R_{BCS}$ (b) for bulk Nb and Nb film cavity before and after each annealing. The $R_{res}$ exhibited a significant field dependency, while the $R_{BCS}$ resistance showed minimal field dependence.

## C. Temperature mapping results

During RF testing after Nb film deposition, temperature maps were collected to observe the distribution of power dissipation on the cavity surface. From the 576 sensors, eight sensors were excluded from the analysis due to readings beyond the acceptable range caused by a calibration issue. At low fields, many sensor readings exhibited small fluctuations of approximately ±10 μK, which are attributed to mechanical noise and intrinsic properties of the measurement hardware, rather than the RF field itself. The accelerating gradient was gradually increased in small steps, resulting in minimal heating by approximately 100 μK at an accelerating gradient of 4 MV/m. At the quench field, a temperature spike was observed in the lower elliptical part of the cavity, as depicted in Figure 4. Annealing at 340 °C resulted in a 30° shift in the quench area within the same cavity half-cell. Following all vacuum furnace annealings, the quench area aligned with that observed in the Nb film cavity that underwent no heat treatment other than for 900 °C annealing for 6 hours. This annealing condition led to a significant shift of the quench area towards the upper elliptical portion of the cavity, as illustrated in Figure 4. Temperature change of specifically selected five areas UHC1, UHC2, LHC1, LHC2, and EQRT will be analyzed in the discussion section.

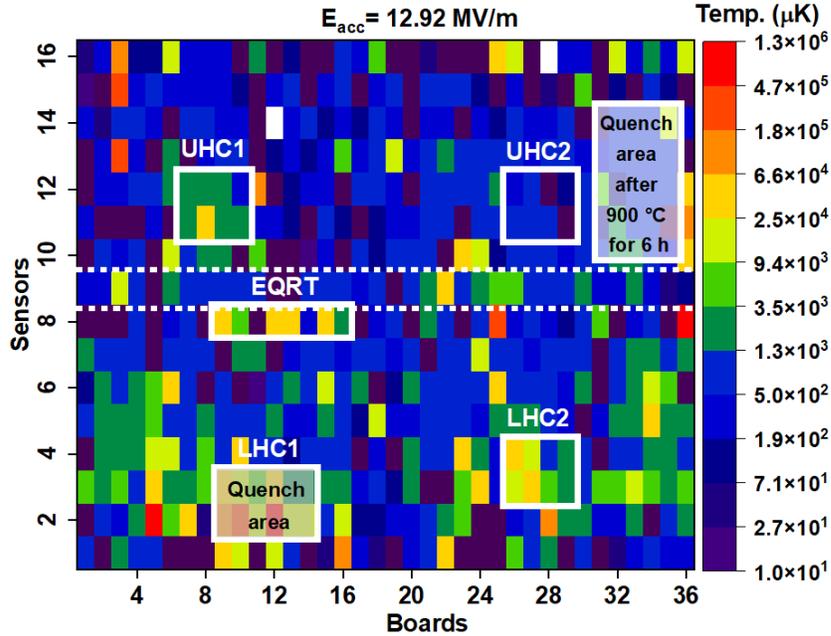

Figure 4. Temperature map of the Nb film cavity at $E_{acc} \cong 12.9$ MV/m. The power dissipation across the surface of the Nb film was not uniform, with certain localized areas exhibiting significantly higher temperatures than others.

## IV. DISCUSSION

In the present experiments, a drastic change in the field dependence of the $R_s$ was observed after the bulk Nb cavity was coated with Nb film. Prior to Nb film deposition, bulk Nb cavity exhibited a typical field dependence expected from an electropolished bulk Nb cavity. After Nb film deposition, the $R_s$ of the Nb film cavity exhibited a pronounced dependence on the applied field prior to the heat treatments. The $R_{BCS}$ was exceptionally low at about 3-4 n$\Omega$ and exhibited weak dependence on applied field. The losses were dominated by the $R_{res}$, which displayed a strong dependence on the RF field. After each higher temperature annealing, a significant improvement in the maximum field of the Nb film cavity was observed and the $R_s$ became less field dependent. With the high temperature annealings, $R_{BCS}$ increased but remained weakly field-dependent, while the $R_{res}$ decreased and displayed reduced field dependence. It is important to look at the changes in $R_{res}$ and $R_{BCS}$ and discuss what factors are contributing to these changes in the $R_s$ of the Nb film cavity.

Following film deposition, the $R_{BCS}$ exhibited a notable reduction from 10 to 3 n$\Omega$, as presented in Figure 3. Although *in-situ* annealing had a minimal effect on $R_{BCS}$, subsequent vacuum furnace annealings at 600 °C and 800 °C increased $R_{BCS}$, to 6 and 8 n$\Omega$, respectively, bringing it closer to the $R_{BCS}$ of bulk Nb cavity. The significant improvement in $R_{BCS}$ after Nb film deposition can be attributed to the lower RRR of the sputtered films [35]. High RRR bulk Nb cavities typically exhibit a long mean free path (mfp ~ 1000 nm) after the standard chemical etching, which is not optimal for $R_{BCS}$, as the minimum in BCS surface resistance as a function of the mfp is found when the mfp is close to the coherence length ($\xi_0$ ~ 38 nm). In the case of Nb film cavities, RRR is typically 10-30 for DC or HiPIMS sputtered films [4, 6, 28, 35], which corresponds to mfp of 27-90 nm. This proximity of the mfp to the $\xi_0$ results in a BCS surface resistance is close to the minimum [1, 2], as observed in this experiment, a threefold reduction in $R_{BCS}$ was noted when comparing bulk Nb to Nb film. Both *in-situ* and vacuum furnace annealings are known to change impurity distribution and its concentration within the RF penetration layer [1, 2].

*In-situ* annealing at 340 °C for 1 hour resulted in the dissolution of the surface oxide and an increase in surface carbon and nitrogen concentrations but led to only slight changes in impurity concentrations within the film, as shown in our previous work [35], yielding no significant improvement in $Q_0$ or its strong field dependence. These findings suggest that the surface oxide is not the primary contributor to the medium-field Q-slope. On the other hand, vacuum furnace annealing at elevated temperatures produced substantial changes in impurity concentrations, as demonstrated by SIMS depth profiling in the same study and led to notable improvements in both the mfp with RRR increasing from 23 to 50 for the Nb film cavity. SIMS analysis indicated that the oxygen concentration within the film remained largely unchanged after annealing treatments. This stability is consistent with the system's geometry—a 6 μm-thick Nb film on a 3 mm-thick Nb substrate—which effectively behaves as an infinite medium for oxygen diffusion. Under the high-vacuum conditions used during furnace annealing, and given the limited oxygen supply from the native oxide layer, oxygen redistribution within the film was minimal. However, following the high-temperature treatments, exposure of the cavity to air led to the reformation of a native oxide layer on the film surface. In contrast to oxygen, hydrogen concentrations decreased significantly as shown in Figure 6, along with a moderate reduction in nitrogen levels compared to those observed after *in-situ* annealing [35]. These changes contributed to increase in mfp and the $R_{BCS}$, bringing the latter closer to values typically measured in bulk Nb cavities prior to film deposition.

While $R_{BCS}$ and its changes with annealings in the present experiment can be adequately explained by the changes in the mfp, it is challenging to explain significant changes in $R_{res}$ with the heat treatments. To delineate the field dependence of the $R_{res}$, we developed simplified version of the model of *Visentin et al.* [31]. A quadratic relationship between $R_{res}$ and the accelerating electric field $E_{acc}$ was established for medium field levels, specifically for $E_{acc} > 4$ MV/m, as expressed in Equation 1. This relationship is illustrated in Figure 5.

$$R_{res}(\alpha) = R_m + \alpha \cdot E_{acc}^2 \qquad (1)$$

where $R_m$ is the residual resistance at $E_{acc} \cong 4$ MV/m and $\alpha$ is the free parameter that represents the medium-field Q-slope. The resulting fitting parameters are presented in Table 1, where $R^2$ is the coefficient of determination, further support our model.

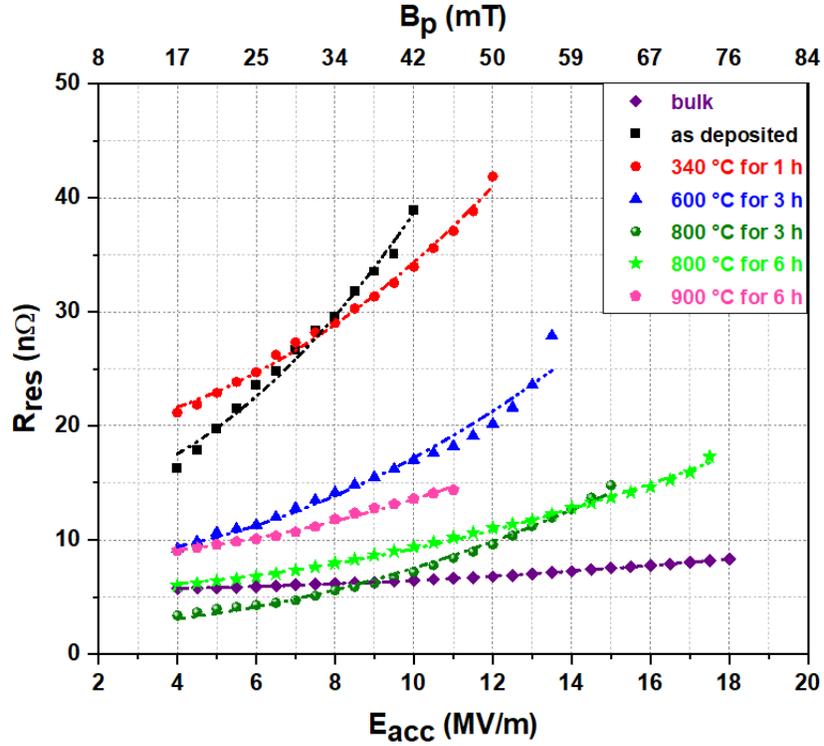

Figure 5. Residual resistance of bulk Nb and Nb film cavity before and after annealing treatments fit with Equation 1. Field dependency of the results indicate that the field dependency of $R_{res}$ decreased as a result of the annealing procedures. This reduction in field dependency suggests that the heat treatments effectively improved the material properties.

Table 1. Fitting parameters of the $R_{res}$ with respect to $E_{acc}$.

| Condition | $R_m$ (n$\Omega$) | $\alpha$ (×10$^{-3}$) (n$\Omega$/(MV/m)$^2$) | $R^2$ |
|---|---|---|---|
| Bulk Nb | 5.7 | 9.2 | 0.98 |
| As deposited | 16.2 | 250.1 | 0.94 |
| After annealing at 340 °C for 1 hour | 21.1 | 153.2 | 0.92 |
| After annealing at 600 °C for 3 hours | 9.1 | 109.4 | 0.89 |
| After annealing at 800 °C for 3 hours | 3.4 | 71.3 | 0.92 |
| After annealing at 800 °C for 6 hours | 6.2 | 41.7 | 0.95 |
| After annealing at 900 °C for 6 hours | 9.0 | 50.5 | 0.96 |

The analysis reveals that the $R_m$ of the Nb cavity after Nb film deposition increased by approximately a factor of three over that of the bulk Nb cavity. Normal conducting inclusions, magnetic fluxoids, sub-oxides are known to increase the $R_{res}$ of SRF cavities. One of the differences in the film morphology is the fine grain size of deposited film as compared to bulk Nb. The increased granularity leads to a higher concertation of grain boundaries, dislocations and weak link between grains [37-43].

After subjecting the Nb film cavity to *in-situ* annealing at 340 °C for 1 hour, $R_m$ increased from 16.2 to 21.1 nΩ. This annealing is known to cause dissolution of the $Nb_2O_5$ layer into $NbO_2$ and NbO, with the excess of oxygen diffusing into the Nb [2, 44]. The *in-situ* annealing led to an increased formation of suboxides, contributing to a rise in the $R_m$ [44]. This increase in $R_m$ can also be attributed to elevated concentrations of carbon and nitrogen at the film surface, as previously reported [35].

Subsequent vacuum furnace annealing at elevated temperatures led to a significant reduction in niobium hydrides and their associated field-dependent $R_{res}$. One of the primary reasons for this reduction is the effective degassing of hydrogen, as demonstrated by applying the standard procedure of heat treatment of bulk Nb cavities at 800 °C to lower their hydrogen content [45-48]. Niobium possesses a strong affinity for hydrogen, resulting in substantial absorption of hydrogen during processing, such as chemical etching when the natural oxide layer is absent [1, 47]. Upon cooling to cryogenic temperatures, absorbed hydrogen can bond to niobium, forming niobium hydrides that significantly increase $R_{res}$ and its field dependence, thus contributing to the phenomenon known as Q-disease [45-48]. To investigate this effect, SIMS depth profiling was performed and revealed a high hydrogen concentration within the Nb film in the as-deposited state, as shown in Figure 6. For each condition, three separate samples were annealed, and three distinct areas were analyzed on each sample, providing a more robust and representative assessment of the hydrogen content. Following high-temperature annealing, formed niobium hydrides decreased significantly compared to as-deposited samples and for samples annealed at 340 °C for 1 hour. This reduction in hydrogen correlates with the observed improvement in $Q_0$ and a decrease in loss mechanisms. Although the hydrogen concentration is expected to homogenize across the cavity due to its long diffusion length at room temperature leading to a uniform anomalous Q-slope [1, 2], temperature mapping analyses indicated that the Q-slope also originated from localized areas.

This localized nature of the anomalous losses suggests that additional loss mechanisms beyond hydrides may also be contributing to the observed Q-slope issue.

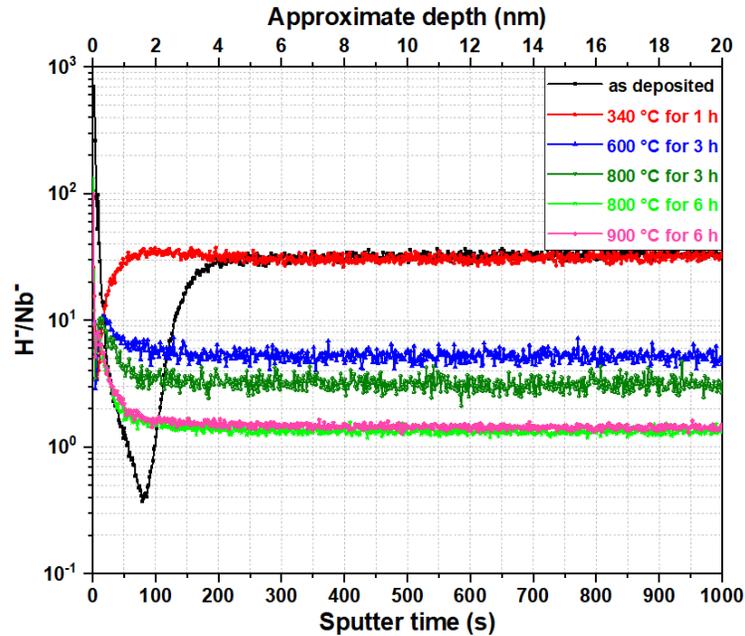

Figure 6. Relative concentration of hydrogen as a function of depth before and after the annealing procedures. The results demonstrate a tenfold decrease in hydrogen concentration within the Nb film following high-temperature annealing treatments.

Furthermore, high local misorientation angles are correlated with temperature increases when an RF field is applied to the SRF cavity [49]. This localized temperature increase contributes to an increase in $R_s$. The local misorientation angle distributions (LMAD), obtained from multiple 15 × 15 μm regions in the EBSD maps of each sample and normalized to the total scanned area, are presented in Figures 7 and 8. Following each heat treatment, EBSD maps exhibited (101) grain orientation (z-axis: normal to the surface) is preferential and LMAD peaks shifted towards lower misorientation angles. Peak position of the LMAD for both the as-deposited samples and those subjected to 340 °C annealing for 1 hour were approximately 0.25 °. However, subsequent heat treatments resulted in further shifts to lower misorientation angles. For instance, annealing at 600 °C and 800 °C for 3 hours reduced the peak positions to 0.19 ° and 0.16 °, respectively. Additionally, annealing at 800 °C for 6 hours further shifted the peak position to 0.10 °. In contrast, 900 °C annealing for 6 hours did not result in a remarkable change in the peak position or the LMAD, indicating no further relaxation and recrystallisation of the grains [50, 51]. The observed changes in local misorientation angles are expected to reduce the values of $\alpha$ and $R_{res}$ and reduce filed dependency of the $R_{res}$. This trend underscores the importance of controlling misorientation during the film deposition and through heat treatments to enhance the performance of Nb film cavities.

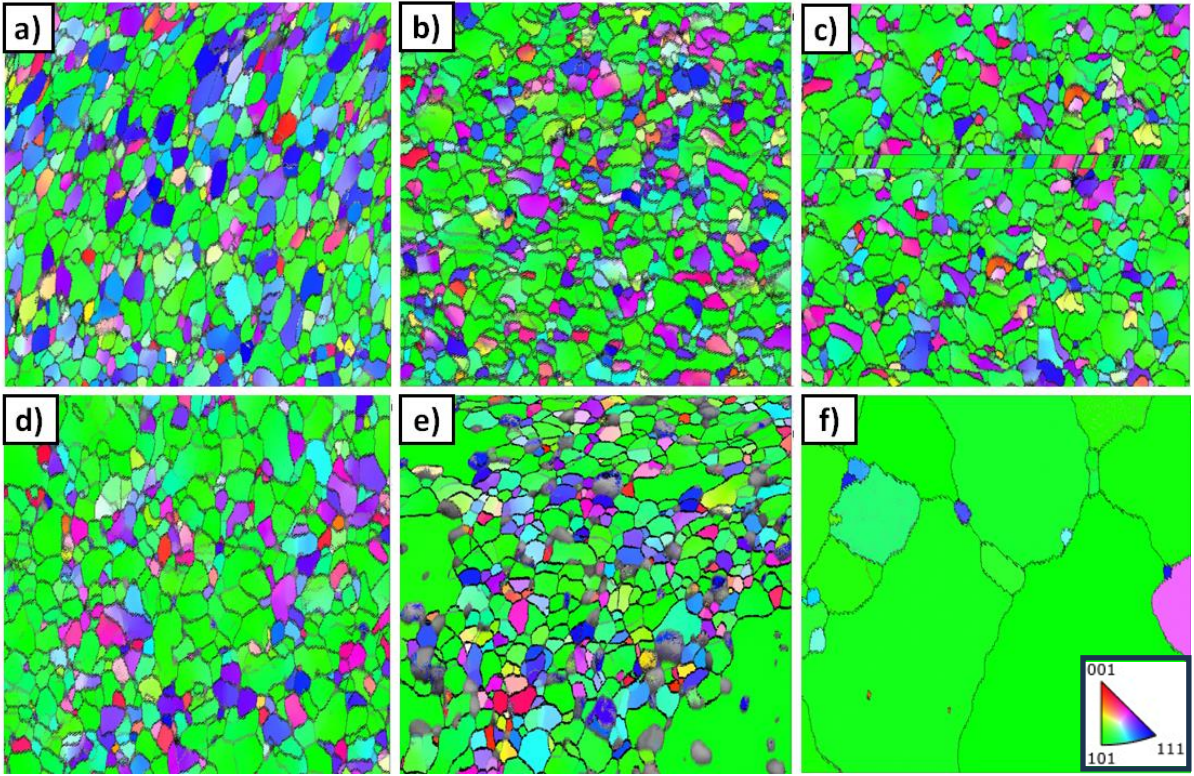

Figure 7. EBSD map of the Nb film before and after heat treatments (a) as-deposited, (b) after annealing at 340 °C for 1 hour, (c) at 600 °C for 3 hours, (d) at 800 °C for 3 hours, (e) after at 800 °C for 6 hours, (f) at 900 °C for 6 hours. The legend is an inverse pole figure whose color code is for each grain orientation. After annealing at 800 °C, significant grain growth was observed, accompanied by a pronounced dominance of the (110) crystallographic orientation.

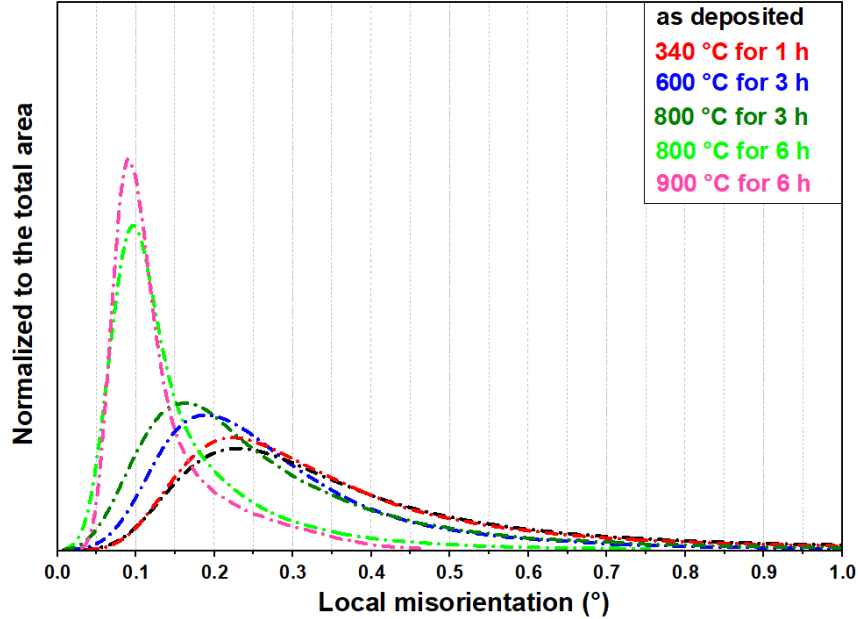

Figure 8. Local misorientation angle distribution of the 15 μm × 15 μm area of the Nb film before and after heat treatments. Following each annealing procedure, the peak position of the LMAD shifted towards lower angles. The full width at half maximum of the distribution became narrower, reflecting a reduction in misorientation variability.

The Nb film cavity experienced localized hot spot areas in the lower elliptical section of the cavity and quenched in one of these areas, as shown in Figure 3. Although the surface roughness is higher in the equator region due to the welding process used during cavity fabrication, the quench location was identified in the lower half-cell, away from the equator. The finding points to defects as the source of heating, consistent with the temperature mapping analysis in [52]. Further investigation will be required to ascertain the nature of quench defect within the cavity or evaluate the quality of coating at the quench site.

Non-uniformity in losses on the cavity surface was further evaluated with the analysis of the temperature maps. Temperature maps revealed distinct heating patterns on the cavity surface, with one half cell of the cavity surface exhibiting higher temperatures compared to the other half cell, as shown in Figure 4. To better understand the field dependence of the surface resistance of the cavity surface, five different areas on the cavity surface were analyzed. These areas are marked with white rectangles in Figure 4.

The power dissipated into the cavity wall can be related to the average surface temperature rise via Equation (2):

$$P_{diss}^{tot} = \sum_{i=1}^{576} c_i \cdot \Delta T_i \cdot A = k \cdot T_{av} \quad (2)$$

where $c_i$ is the efficiency coefficient of each sensor, $\Delta T_i$ is the temperature increase, $A$ is an effective area of each temperature sensor, $k$ is the effective coefficient, and $T_{av}$ is the average temperature increase of the sensors. Equation (2) establishes a relationship between RF power dissipation and the average temperature increase, measured with the temperature sensors. Subsequently, the $Q_0$ of the cavity can be determined as follows:

$$Q_0 = \frac{2\pi f \cdot U}{P_{diss}^{tot}} = \frac{2\pi f \cdot U}{k \cdot T_{av}} \quad (3)$$

where $f$ is the frequency, and $U$ is the stored energy in the cavity. This analysis is applied to the selected areas on the cavity surface after annealing at 600 °C for 3 hours, shown in Figure 4, to evaluate $Q_0$ and RF losses of each region.

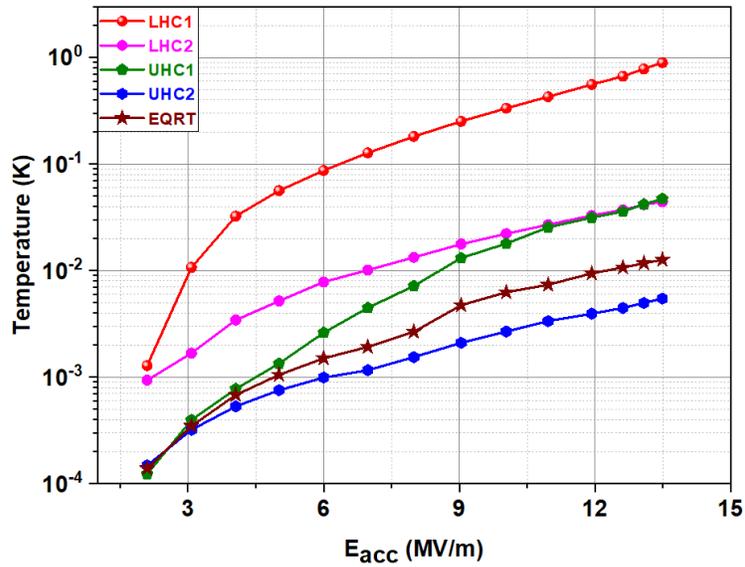

Figure 9. Average temperatures of the areas selected in Figure 4. The average temperature of the LHC1 (quench) area is found to be higher than that of the other areas.

Figure 9 illustrates the average temperatures for each of the selected areas. These temperatures serve as the basis for calculating the equivalent $Q_0$ as a function of field, as illustrated in Figure 10. The analysis reveals that the quench area, LHC1, and UHC1 dominate the cavity surface resistance and its field dependence, while other areas have weakly field dependent surface resistance, e.g., LHC2, exhibiting like bulk Nb RF surface resistance. The anomalous field-dependent surface resistance, as observed in the integral measure $Q_0$, is therefore determined by a few surface regions.

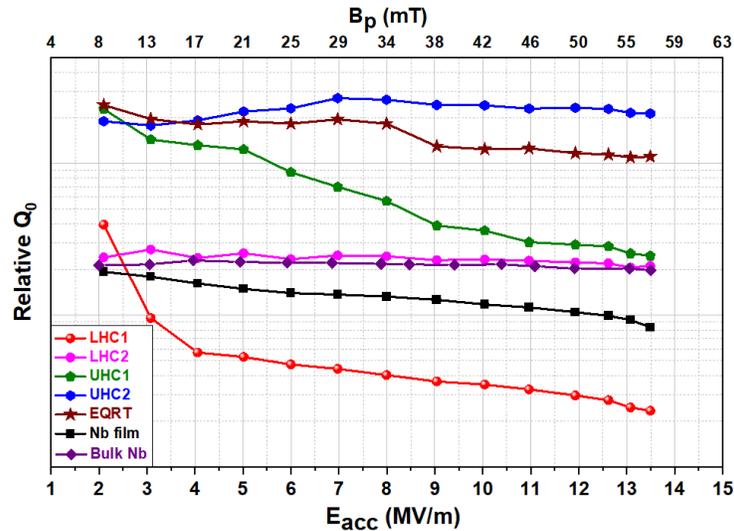

Figure 10. $Q_0$ as a function of field calculated from the average temperatures of temperature sensors selected from characteristic areas on the cavity surface. The $Q_0$ values for the LHC1 and UHC1 regions exhibited a strong dependence on the applied field, whereas other regions demonstrated negligible field dependence.

Improvements of the HiPIMS film deposition technique for Nb film deposition over DC sputtering are high adhesion, low roughness, and the dense columnar structure [26-28, 53]. HiPIMS Nb film deposition technique continues to be optimized to approach bulk-like material properties and tailor them to application on 3D surfaces. Some of the open questions are the impact of the grain size, the contribution of porosities at grain boundaries, and the contribution of voids between substrate and film to the $R_s$, which should not be assumed negligible [27, 28, 53, 54]. Previous research demonstrated that using DC bias HiPIMS reduced voids between the substrate and film by 87% compared to DC magnetron sputtering [53]. However, some voids persist due to substrate surface properties and the native oxide layer. These voids can increase the strain in the lattice, and lattice defects at grain boundaries, like vacancies or dislocations, and may contribute to an increase in $R_s$ [53, 54]. These issues could be further aggravated by the defects on the substrate such as pits or chemical residue, which will lead to higher strain in deposited film. High temperature annealing effectively dissolves the oxide layer [2] and was suggested as one of the approaches to improve Nb/Cu interface thermal boundary [17]. High temperature annealing also reduces lattice defects and porosities at grain boundaries [55, 56].

## V. CONCLUSION

Our analysis focused on the RF performance of a 1.3 GHz single-cell Nb cavity coated with a Nb film using DC biased HiPIMS. At high fields, the performance of the Nb film cavity was limited by quench at $E_{acc} \cong 10$ MV/m and $R_s$ exhibited a significant dependency on the applied accelerating gradient. Oxide free surface after the *in-situ* annealing at 340 °C for 1 hour led to a modest improvement in the quench field ($E_{acc} \cong 12.5$ MV/m), without significantly altering the $R_{BCS}$ or $R_s$ the field dependence. Further vacuum furnace annealing at 600 °C and 800 °C for 3 hours resulted in reductions in $R_s$ and its field dependence, increasing the quench field to $E_{acc} \cong 13.5$ MV/m and $E_{acc} \cong 15.3$ MV/m, respectively. A subsequent annealing treatment at 800 °C for 6 hours further improved the quench field to record high $E_{acc} \cong 17.5$ MV/m for Nb film cavities. However, increasing the temperature to 900 °C induced the Q-switching phenomenon. The consistent reduction in field-dependent losses with each heat treatment correlated with a decrease in hydrides and lattice misorientation, as confirmed by material characterization. TMAP further revealed several localized hot spots contributing to the medium-field Q-slope. These findings suggest that hydrides, high local misorientation, and structural defects are key contributors to the field-dependent losses in Nb film cavities.

Having identified the principal material limitations, future efforts should prioritize the optimization of Nb film deposition on Cu substrates. Specifically, reducing hydrogen content, minimizing microstrain, and improving film uniformity while avoiding defect formation will be critical to enhancing the performance and mitigating field-dependent losses in Nb/Cu SRF cavities.

## ACKNOWLEDGEMENT

This manuscript has been authored by Fermi Forward Discovery Group, LLC under Contract No. 89243024CSC000002 with the U.S. Department of Energy, Office of Science, Office of High Energy Physics. Additionally, this material is based upon work supported by the U.S. Department of Energy, Office of Science, National Quantum Information Science Research Centers, Superconducting Quantum Materials and Systems Center (SQMS) under contract number DE-AC02-07CH11359.